\documentclass[12pt,preprint]{aastex}

\shortauthors{Gordon}

\newcommand\ee{\end{equation}}
\newcommand\be{\begin{equation}}
\newcommand\eea{\end{eqnarray}}
\newcommand\bea{\begin{eqnarray}}

\newcommand\fig[1]{Figure~#1}

\newcommand\eq[1]{Eq.~(\ref{#1})}

\renewcommand\d[1]{{\bf \hat{#1}}}
\newcommand\chieff{\chi^2_{\rm eff}}
\newcommand{\beq}{\begin{equation}}
\newcommand{\eeq}{\end{equation}}

\begin{document}

\title{ Broken Isotropy from a Linear Modulation of the Primordial
  Perturbations } 
\author{Christopher Gordon\altaffilmark{1,2}}
\altaffiltext{1}{
Astrophysics, University of Oxford,
Denys Wilkinson Building, Keble Road, Oxford OX1 3RH, UK}
\altaffiltext{2}{ KICP and EFI, %
University of Chicago, Chicago, IL 60637, USA}
\email{cxg@astro.ox.ac.uk}
\slugcomment{Draft: \today}

\begin{abstract}
A linear modulation of the primordial perturbations is proposed as an
explanation for the observed asymmetry between the northern and southern
 hemispheres of the Wilkinson Microwave Anisotropy Probe ({\em
   WMAP\/}) 
data. 
A cut sky, reduced resolution third
 year ``Internal Linear Combination'' ({\em ILC\/}) map was used to estimate
 the modulation parameters. A 
 foreground template and a modulated plus unmodulated monopole and
 dipole were projected out of the likelihood. The effective chi squared
 was 
reduced by nine for three extra parameters. The mean galactic colatitude
 and longitude, of the modulation, with 68\%, 95\% and 99.7\% confidence intervals were
$56^{+17 +36 +65}_{-17 -35 -51}$ and $63^{+28 +59 +105}_{-26  -58
   -213}$. The mean percentage change of the variance, across the
 pole's of the  modulation,
was $62^{+18 +35 +57}_{-18 -35 -47} $. Implications of these
 results and possible generating mechanisms are discussed.
\end{abstract}
\keywords{cosmic microwave background, early universe, large-scale
  structure of universe}

\maketitle

\section{Introduction}
A fundamental assumption of cosmology is that the Universe
is isotropic. 
This was confirmed, for the mean temperature of
the cosmic microwave background ({\em CMB\/}), by the  {\em FIRAS\/}
experiment on the {\em COBE\/} satellite \citep{wright92,bennett96}.
However, the higher precision, measurements from the 
Wilkinson Microwave Anisotropy Probe
({\em WMAP}) satellite 
\citep{bennett03,hinshaw06,jarosik06,page06,spergel06}, have 
 an anomalously asymmetric distribution, in the temperature 
fluctuation statistics,
between the northern  and southern hemispheres of the sky
\citep{erihanbangorlil03,hanbangor04,
vielva03,park03,cophutsta03,hansen04,larwan04,crumarvie04,lanmag04,hansen04a,bernui05,bernui06}. On scales greater than
about $5^\circ$, 
the variance of the {\em CMB\/} temperature
fluctuations is anomalously higher in the southern hemisphere, in
both galactic and ecliptic coordinates,
compared to the northern hemisphere \citep{erihanbangorlil03,hanbangor04}.
 This asymmetry  also appears
in  higher order statistics \citep{vielva03,park03,cophutsta03,hansen04,larwan04,crumarvie04,lanmag04,hansen04a,bernui05,bernui06}.

In a spherical harmonic representation, scales ranging from  $\ell=2$ to
$\ell=40$ were found to be asymmetric. When optimized over
direction, only 0.3\% of isotropic
simulations   were  found to produce   higher  levels of 
asymmetry \citep{hanbangor04}.

The result is not sensitive to the
frequency band of the {\em CMB\/} \citep{hanbangor04} and a similar pattern
(at lower significance) is seen in {\em COBE\/} \citep{hanbangor04}. This
argues against a 
foreground or systematics explanation.

Although a simple single field inflation model would
give isotropically distributed perturbations, this is not necessarily
the case in multi-field models \citep{linmuk05}. Thus, if it
can be shown that the {\em CMB\/} fluctuations are not isotropic, it may be an
indication that inflation was a multi-field process.

The layout of the paper is as follows:
In Sec.~\ref{sec:modulation}, a linearly modulated primordial power
spectrum is proposed as the source of the observed isotropy breaking.
Then, in Sec.~\ref{sec:likana}, a method of evaluating the linear
modulation parameters is outlined. The constraints are given in
Sec.~\ref{sec:results} and their implications and relation to other
results are discussed in Sec.~\ref{sec:discussion}.

\section{Linear Modulation}
\label{sec:modulation}
An isotropy-breaking mechanism may be parameterized as
\citep{pruuzaberbru04,gorhuhut05,spergel06}
\beq 
\delta T(\d{n})=\delta T_{\rm iso}(\d{n})(1-f(\d{n})) 
\label{Tmod}
\eeq 
where $\delta T$ is the observed {\em CMB\/}
temperature perturbations, $\delta T_{\rm iso}$ are the underlying
isotropically distributed temperature perturbations and $\d{n}$ is the
direction of observation. 
An isotropic
distribution of perturbations is recovered when $f=0$.

\citet{spergel06} parameterized $f$ as
\beq
f(\d{n})
=\sum_{\ell=1}^j \sum_{m=-\ell}^\ell f_{\ell m} Y_{\ell m}(\d{n})
\label{fsph}
\eeq  
where $Y_{\ell m}$ are spherical harmonics and $j=1$ and $j=2$ were
tried. 
As $f$ will be a real function, the
condition $f_{\ell m} = (-1)^m f_{\ell -m}^*$ is required, where the asterisk
indicates the complex conjugate. 
The
 effective chi squared improvement, $\Delta \chieff\equiv 
-2\Delta\log {\cal L}$ is only -3 for the $j=1$ case and only -8 for the $j=2$ case,
  where $\cal L$ is the likelihood \citep{spergel06}.

The dipolar modulation ($j=1$) case
has the potential to  explain the lack of isotropy between
  two hemispheres, as it will reduce the
  variance of the perturbations in one hemisphere and increase it in
  the other.

In this article, an underlying spatial model for a dipolar modulation is
investigated.
On large scales ($\ell < 30$)  the main contribution \citep{husug94}
to the {\em CMB\/} perturbations is the Sachs Wolfe effect \citep{sw}
\beq
{ \delta T(\d{n}) \over T} \approx {1 \over 3} \Phi(\d{n})
\eeq
where $\Phi$ is the curvature perturbation, in the Newtonian
gauge \citep{bardeen}, evaluated at last scattering.\footnote{This is the
  surface where the Universe becomes effectively transparent to
  photons and occurs at a redshift of about 1000.}
The left hand side is the measured temperature fluctuation divided by
the average measured temperature.

It follows that on large scales and at last scattering
\beq
\Phi=(1-f)\Phi_{\rm iso}\, .
\eeq 
A dipolar
modulation of the last scattering surface could result from a spatially
linear modulation
\beq
f(\d{n}) = {\bf w} \cdot \d{n}
\label{fspatial}
\eeq 
where ${\bf w}\equiv (w_x,w_y,w_z)$ is the gradient of the modulating
function and the dot indicates a dot product.

The spherical harmonic  representation is related to
a Cartesian representation by
\beq
\left[f_{10},{\rm Re}(f_{11}),{\rm Im}(f_{11})\right]
=\sqrt{\pi\over 3}\left[2 w_z,-\sqrt{2}w_x,\sqrt{2}w_y\right]
\eeq
As they are linearly related, a uniform prior on one translates into a
uniform prior on the other.
 
\section{Likelihood Analysis}
\label{sec:likana}
The likelihood  is assumed to follow a multivariate
Gaussian distribution
\beq
{\cal L} \propto |{\bf C}|^{-1/2} \exp\left(-{1\over
    2}  {\bf T}^{\rm T}
{\bf C}^{-1}  {\bf T} \right)
\eeq
where ${\bf T}$ is a vector of the temperature measurements in
unmasked areas of the {\em CMB\/} pixelized map. Each element of the covariance matrix, ${\bf
  C}$,
is evaluated by
\beq
C(\d{n},\d{m})=(1-f(\d{n})) C_{\rm iso}(\d{n},\d{m})
(1-f(\d{m}))+\lambda C_{\rm marg}(\d{n},\d{m})
\eeq
where $C(\d{n},\d{m})$ corresponds to the covariance between pixels
at position $\d{n}$ and $\d{m}$.
The instrumental noise is negligible on large scales
\citep{jarosik06,hinshaw06} and so is not included.
 The underlying isotropic covariance matrix, between pixels in
 directions $\d{n}$ and $\d{m}$,
  can be decomposed as 
 \beq
 C_{\rm iso}(\d{n},\d{m}) =\sum_{\ell} C_{\ell} w_{\ell}^2  P_{\ell}(\d{n}\cdot\d{m})
 (2\ell+1)/(4\pi)\,.
 \label{coviso}
 \eeq
 where $C_{\ell}$ is the angular power spectrum, $P_\ell$ is the Legendre polynomial of order $\ell$ and
 $w_\ell$ is the effective window function of the smoothed map evaluated
 at a low pixel resolution. Modulating the temperature perturbations
 leads to a transformed covariance matrix of the form
 \beq
 C(\d{n},\d{m})=(1-f(\d{n})) C_{\rm iso}(\d{n},\d{m})
 (1-f(\d{m})) \, .
 \label{nonisoC}
\eeq
A marginalization term \citep{
tegmark97,bonjafkno97,sloselmak04,slosel04,
hinshaw06} 
for foregrounds and a modulated and unmodulated
monopole and dipole was also  added
\beq
C_{\rm marg}= {1\over 2}(C_0+C_1)+{1\over 2}(1-f)(C_0+C_1)(1-f)+C_{\rm
  foregrounds}\,.
\label{Cmarg}
\eeq
The unmodulated monopole and dipole terms are needed to account for any
residual effects of
the   background temperature and peculiar motion
of the observer \citep{sw}. When $\lambda$ is made sufficiently large,
the likelihood becomes insensitive to any terms included in  $C_{\rm
  marg}$ \citep{
tegmark97,bonjafkno97,sloselmak04,slosel04,hinshaw06}.   

Third year {\em WMAP\/} data were used\footnote{Obtained from
  http://lambda.gsfc.nasa.gov/product/map/current/} and the 
preprocessing followed was the  same 
as in the {\em WMAP\/} analysis \citep{hinshaw06,spergel06} for
  their large scale likelihood evaluation. 
The $N_{\rm side}=512$
 ``Internal Linear Combination''
  ({\em ILC\/}) map\footnote{number of pixels =$12N_{\rm side}^2$.}
was masked with the Kp2 mask and smoothed with a  
$7.3^\circ$ (FWHM) Gaussian smoothing function. It was then degraded
  to $N_{\rm side}=8$ 
using the {\em HEALPix}\footnote{http://healpix.jpl.nasa.gov} software
package \citep{gorski05}.  
The Kp2 mask, consisting of
zeros and ones, was also
degraded to $N_{\rm side}=8$  and any pixels with values larger than
  0.5 were set to one, else they were set to zero. The smoothed degraded {\em ILC\/} map was then
remasked with the degraded Kp2 mask. The foreground template was taken
  to be the difference between the raw V-band map and the {\em ILC\/}
  map. The foreground marginalization term in \eq{Cmarg} was
  set equal to the
  outer product of the foreground template with itself
  \citep{
tegmark97,bonjafkno97,sloselmak04,slosel04,hinshaw06}.

The $C_{\ell}$ values were treated as free parameters for $\ell =2$ to
10. For $\ell=11$ to 32, the unmodulated maximum likelihood values
were used. The smoothing and degrading of the data make the likelihood
insensitive to  $\ell>32$.

The likelihood, for the modulated model, was numerically maximized
using a quasi-Newton method.

Marginalized distributions of the parameters were obtained using
the Metropolis algorithm. 
After an initial burn in run, a proposal covariance matrix was constructed from
20000 samples. This was used to generate three additional sets of
20000 samples, each with a different starting value chosen from a burned
in chain. The Gelman and Rubin test were used to check
convergence and then the 60000 samples were used to evaluate the
marginal probability distributions of the parameters. All priors were
taken to be uniform. The upper and lower limit  of each 
confidence interval was chosen so as to exclude the same number of
samples above and below the interval.

\section{Results}
\label{sec:results}
\begin{figure}
\plotone{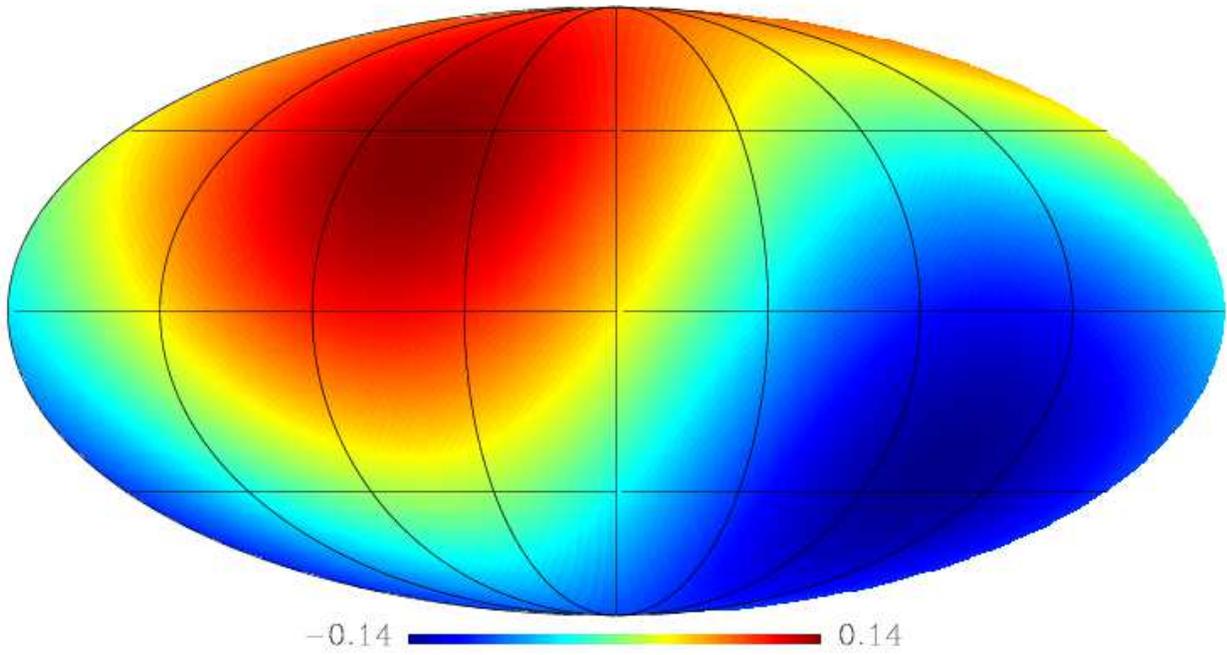}
\caption{\footnotesize
Maximum likelihood linear modulation function projected on the last scattering
surface. High values indicate
suppression of the primordial perturbations and low values indicate
enhancement. 
  }
\label{maxlik}
\end{figure}

\begin{figure}
\includegraphics[angle=90,scale=0.51]{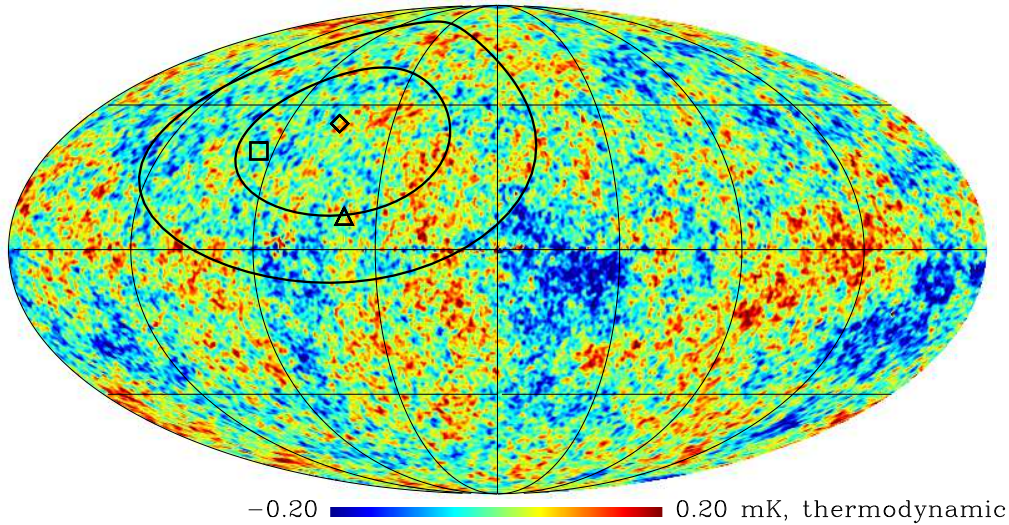}
\caption{\footnotesize
The contours enclosing 68\% and 95\% of the Monte Carlo samples. The
maximum likelihood ($\diamond$)  Galactic co-latitude and  longitude is
$(51^\circ,68^\circ)$.  One of the most
asymmetric directions found by \citet{hanbangor04}
($\triangle$) is ($80^\circ,57^\circ$). The north ecliptic ($\square$)
is at ($60^\circ,96^\circ$). This is plotted over the third year $N_{\rm side}=
512$ {\em ILC\/} map \citep{hinshaw06}. 
  }
\label{fig:results}
\end{figure}

\begin{figure}
\plotone{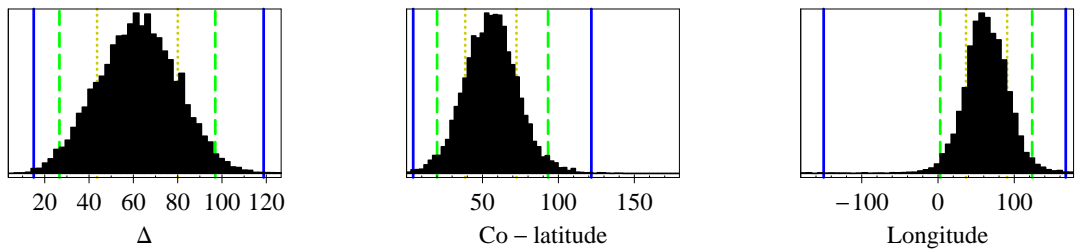}
\caption{\footnotesize
Marginalized distributions of the percentage change in the variance
across the poles ($\Delta$) and the galactic coordinates of the
direction of the modulation. The 68\%, 95\% and 99.7\% confidence
intervals are also shown.
  }
\label{fig:pdfs}
\end{figure}

The improvement, in the likelihood, compared to an unmodulated model was
$\Delta \chi^2_{\rm eff}  = -9$ with three extra
parameters $(w_1,w_2,w_3)$. A plot of the maximum likelihood
modulation function $f$,
 is shown  in \fig{\ref{maxlik}}.
The marginalized distributions of $C_2$ to $C_{10}$ were found not to be
significantly different from those in an unmodulated model
\citep{hinshaw06}. The weight vector samples $(w_x,w_y,w_z)$ were
converted into Galactic co-latitude
(varying between $0^\circ,180^\circ$), longitude (varying between $-180^\circ$ and 
$180^\circ$)    and 
percentage change of temperature variance in the direction of symmetry breaking
 $(\Delta \equiv 200 (w_x^2+w_y^2+w_z^2))$. The two
dimensional confidence intervals for the co-latitude and longitude are
shown in \fig{\ref{fig:results}}. 
The marginalized one dimensional
distributions are shown in \fig{\ref{fig:pdfs}}.
Only $0.2\%$ of the samples were more than $90^\circ$ from the maximum
likelihood point.
The results are summarized in Table~1.

\begin{deluxetable}{clc}
\tablecaption{Marginalized Statistics}
\tablenum{1}
\tablehead{\colhead{Parameter}  & \colhead{Mean\tablenotemark{a}}& \colhead{{\em ML}\tablenotemark{b}}}
\startdata
$\Delta$ &
$62^{+18 +35 +57}_{-18 -35 -47} $ &57
  \\
Colatitude &
$56^{+17 +36 +65}_{-17 -35 -51}$ & 
51 \\
Longitude &
$63^{+28 +59 +105}_{-26  -58 -213}$&
68
\enddata
\tablenotetext{a}{Including confidence
intervals (68\%,95\%,99.7\%).}
\tablenotetext{b}{Maximum
Likelihood.}
\end{deluxetable}
\section{Discussion and Conclusions}
\label{sec:discussion}
In this article the modulation model investigated by \citet{spergel06}
  has been extended by including a marginalization over the unmodulated
  monopole and dipole. This additional feature is required if the
  apparent isotropy breaking had a primordial origin. Including this
  marginalization improved the $\Delta \chi_{\rm eff}^2$ value from -3
  to -9.

As seen from the confidence
  intervals in Table~1 and \fig{\ref{fig:pdfs}}, the marginalized
  posterior probability of 
$\Delta$ has its maximum  more than three sigma away from the
 unmodulated case ($\Delta=0$). 
The modulated model is also preferred by the  Akaike Information
  Criteria (AIC) \citep{akaike74,magsor06}. It is not preferred by
 the Bayesian Information Criteria (BIC)
 \citep{schwarz78,magsor06}. However, the BIC is an approximation of
 the Bayesian evidence and assumes a prior for the parameters which
 is equivalent to one observation \citep{raftery95}. The Bayesian
  evidence will be
 inversely proportional to the volume of the prior probability
 distribution of the modulation parameters. It may be hard to produce
  a modulation larger than one without effecting the observed dipole.
A reevaluation of the Bayesian evidence is needed to see how 
it depends on the assumed prior.  This could be done using a nested sampling
 algorithm  \citep{mukparlid05} which, unlike the BIC, does not
  require a Gaussian approximation to be made for the posterior distribution.

\citet{spergel06} also evaluated whether there was an additional
 quadrupolar component to the modulation.  This component could
 potentially be useful in explaining the alignment and planarity of
 the quadrupole ($\ell=2$) and octopole ($\ell=3$) seen in the WMAP temperature data
 \citep{oliveira03,schwarz04}. The normal direction of the plane of
 alignment is $(30^\circ,-100^\circ)$. Also, when the coordinate system is
 rotated in the direction of the normal of the $\ell=2,3$ planarity
 there is anomalous power in the $m=3$ component of the $\ell=5$
 multipole \citep{lanmag05}.  \citet{spergel06} found that including a
 dipolar and quadrupolar component to the modulation improved $\Delta
 \chi_{\rm eff}^2$ by only 8 for a total of 8 extra parameters. Higher
 order terms in a spatial modulation could be implemented as terms
 quadratic in the spatial coordinates. Whether these additional terms
 will become significant when an unmodulated monopole and dipole are
 marginalized over will be part of a future investigation.

The effect of marginalization over foregrounds was checked and found
not to play a big role. Similar improvements are
obtained when the foreground corrected V band is used instead of
marginalization. Also, the results are not sensitive to the exact
method of degrading and applying the mask. A Kp2 extended mask
\citep{eriksen06} did not make a significant difference. Including
additional $C_{\ell}$, with $\ell>10$, as parameters to be estimated
(rather than set to their unmodulated {\em ML\/} values), also does not
significantly effect the results.

It is interesting to compare the estimated modulation found in this
article to that of
\citet{hanbangor04}. The 10 most effective axes of
symmetry breaking, for a range of scales, are plotted in their
\fig{24}. A similar area, to the two dimensional confidence
intervals in \fig{\ref{fig:results}}, is covered. Also, their \fig{19} compares the
power spectra in different hemispheres. The range of
values is consistent with the confidence intervals for $\Delta$ in
Table~1 and \fig{\ref{fig:pdfs}}. 

 \citet{pruuzaberbru04}  tested for a dipolar
modulation. However, the largest scale they looked at was
$\ell=20$ to 100 binned. They did not get significant results in that
range. As  the observed modulation only occurs
for $\ell\lesssim 40$ \citep{hanbangor04}, the
$\ell=20$ to $100$ range would not be expected to show significant
modulation when binned.

 \citet{freeman05} propose that the modulation of
$\ell=2$ to $\ell=7$ may be sensitive to any residual unmodulated dipole
component. This is not a concern for the approach taken in this article
as an unmodulated dipole is projected out of the likelihood, see \eq{Cmarg}.

Searches for lack of isotropy using 
 a method based on a bipolar expansion of the two point
correlation function do not detect the north south asymmetry in the
$\ell=2$ to $\ell=40$ range
 \citep{hajsou06,picon05}. 
 The linear modulation model could be used  to understand why the bipolar estimator is insensitive
 to this type of isotropy breaking.

A small scale cut off in the modulation
implies that a linear modulation
of the primordial power spectrum would only apply to 
wave numbers larger than about $4\times 10^{-3}h$~Mpc$^{-1}$. It would be
interesting to evaluate whether this modulation would be detectable in
 future large scale galaxy surveys. However, at a redshift
of one the change in the variance at opposite poles would only be
about four percent, due to the smaller comoving distance.

A number of attempts have been made to explain the asymmetry
in terms of local nonlinear inhomogeneities
\citep{
  moffat05,tomita05a,tomita05,inosil06}. 
It would be interesting to see
if the polarization maps  of the {\em CMB\/} \citep{page06} 
 could be used to  distinguish local effects from a modulation of the
primordial perturbations.

Primordial magnetic fields 
\citep{durkahyat98,chen04,naselsky04}, global topology
\citep{oliveira03,kunz06}, and anisotropic expansion
\citep{berbubkep03,bunberkep05,gumconpel06}  
can also lead to
isotropy breaking. However, in these cases the modulating function is
of higher order than dipolar  and so these mechanisms are better
suited for explaining the alignment between $\ell=2$ and $\ell=3$ and
the high $(\ell,m)=(5,3)$ mode \citep{gorhuhut05}.

An additive template based on a Bianchi $VII_h$ model has been shown
to provide a good fit to the asymmetry \citep{jaffe05}. However, the
model is only empirical as it would require a very open Universe which
is in conflict with many other observations. It is harder for additive
templates to explain the alignment between $\ell=2$ and $\ell=3$
 as this
requires a chance cancellation between an underlying Gaussian field and
a deterministic template \citep{gorhuhut05, lanmag05a}. 

As seen in \fig{\ref{fig:results}}, the maximum likelihood direction of modulation was found to be about $44^\circ$
from the ecliptic north pole. Only about 9\% of the time would two randomly
chosen directions be as close, or closer, together.
This may be an indication that the  modulation
is caused by some systematic effect or foreground.
However, as can be seen from
\fig{\ref{fig:results}}, the confidence intervals, for the direction of
modulation, cover just under half the northern hemisphere. Therefore, the
actual  direction, of modulation may be significantly further away from the 
ecliptic north pole.

Standard single field inflation would produce isotropic perturbations. However, multi-field models, such as in the
curvaton scenario
\citep{
lytwan01,
mollerach90,linmuk96,enqslo01,mortak01}, can produce, what to a particular observer appear to be, non-isotropic perturbations
\citep{linmuk05}. The curvaton mechanism produces a web like structure
in which relatively stable domains are separated by walls of large
nonlinear 
fluctuations. If the mass of the curvaton field is sufficiently small, our
observable Universe could be enclosed within a stable domain.
If we happen to live near one of the walls, of a domain, then the
amplitude of the perturbations will be larger on the side of the
observed Universe closer to the wall \citep{linmuk05}. 
However, if our observed Universe was far enough away from the web walls,
 the very large scale fluctuations would be  linear 
and so isotropy would be unlikely to  appear to be broken \citep{lyth06}.
As the non-isotropic nature only extends to about $\ell=40$
\citep{hanbangor04}, it would be necessary for the inflaton
perturbations to dominate over the curvaton ones for wave numbers larger than
about $4\times 10^{-3}h$~Mpc$^{-1}$. The curvaton produces curvature
perturbations proportional to $V^{1/2}$ \citep{lytwan01}, where $V$ denotes the 
inflaton potential. While the inflaton produces curvature perturbations
proportional to $V^{3/2}/V'$, where $V'$ denotes the slope of the
potential. So if there is a sudden drop in $V$ and $V'$, it is possible
for the non-isotropic curvaton perturbations to dominate for wave
numbers smaller than $4\times 10^{-3}h$~Mpc$^{-1}$ and inflaton
perturbations to dominate for larger wave numbers.   

There are oscillations in the {\em WMAP\/} power spectrum, at
around $\ell=40$, which may be  caused by a change of slope in the inflaton
potential \citep{covi06}. Whether all these elements can be put together
to make a working curvaton model, that fits the data as well as a linear
modulation, 
is still being investigated.

The results presented here provide a parameterization for the
observed asymmetry between different hemispheres of the {\em
WMAP\/} data. 
Having a specific model for the primordial fluctuations will make it
easier to develop new tests for this asymmetry and help determine if
it is a genuine window into new physics at the largest observable scales.

\acknowledgments
 I thank
Olivier Dor\'{e}, Wayne
Hu, Dragan Huterer, Nemanja Kaloper, David Lyth and  Hiranya Peiris for useful 
discussions. 
I was supported by the {\em Beecroft Institute for Particle Astrophysics and
 Cosmology\/} and also by the {\em Kavli Institute for Cosmological Physics\/} under {\em NSF\/}
PHY-0114422.
Some of the results in this paper have been derived using the {\em HEALPix\/}
\citep{gorski05} package. The {\em WMAP\/} data were obtained from the Legacy Archive
for Microwave Background Data Analysis ({\em LAMBDA\/}).



\end{document}